%% file: main.tex
\documentclass[a4paper,USenglish]{lipics-v2021}

\hideLIPIcs  

\usepackage{mathtools}
\usepackage{longtable} 
\usepackage{pdflscape} 
\usepackage[frozencache]{minted}
\usepackage{attachfile}
\usepackage{tikz}

\input{macro}

\title{Reflexive tactics for algebra, revisited}

\author{Kazuhiko Sakaguchi}{Japan}{pi8027@gmail.com}{https://orcid.org/0000-0003-1855-5189}{}

\authorrunning{K. Sakaguchi} 

\Copyright{Kazuhiko Sakaguchi} 



\category{} 

\relatedversion{} 

\supplement{\url{https://github.com/math-comp/algebra-tactics}}

\acknowledgements{We thank Enrico Tassi for his help with \CoqElpi, particularly for adding some
features and fixing performance bottlenecks.
We thank Assia Mahboubi for providing useful examples for identifying issues.
Their comments on early drafts of this paper have also been a great help.}

\nolinenumbers 

\EventEditors{John Q. Open and Joan R. Access}
\EventNoEds{2}
\EventLongTitle{42nd Conference on Very Important Topics (CVIT 2016)}
\EventShortTitle{CVIT 2016}
\EventAcronym{CVIT}
\EventYear{2016}
\EventDate{December 24--27, 2016}
\EventLocation{Little Whinging, United Kingdom}
\EventLogo{}
\SeriesVolume{42}
\ArticleNo{23}

\begin{document}

\maketitle

\begin{abstract}
 Computational reflection allows us to turn verified decision procedures into efficient automated
 reasoning tools in proof assistants.
 The typical applications of such methodology include mathematical structures that have decidable
 theory fragments, e.g., equational theories of commutative rings and lattices.
 However, such existing tools are known not to cooperate with \emph{packed classes}, a methodology
 to define mathematical structures in dependent type theory, that allows for the sharing of
 vocabulary across the inheritance hierarchy. Additionally, such tools do not support homomorphisms
 whose domain and codomain types may differ.
 This paper demonstrates how to implement reflexive tactics that support packed classes and
 homomorphisms.
 As applications of our methodology, we adapt the \coq{ring} and \coq{field} tactics of \Coq{} to
 the commutative ring and field structures of the Mathematical Components library, and apply the
 resulting tactics to the formal proof of the irrationality of $\zeta(3)$ by Chyzak,
 Mahboubi, and Sibut-Pinote, to bring more proof automation.
\end{abstract}

\section{Introduction}
\label{sec:introduction}

Computational reflection~\cite{DBLP:conf/lics/AllenCHA90} makes it possible to replace proof steps
with computations and has been widely used to automate proofs in some proof assistants such as
\Coq~\cite{coqrefman} and \Agda~\cite{DBLP:conf/tphol/BoveDN09}.
For example, we can prove an integer equation $(a - b) - (b - a) = 0$ as follows.
\begin{enumerate}
 \item We obtain a term
       $e \coloneqq \mathrm{Sub}(\mathrm{Sub}(\mathrm{X}_0, \mathrm{X}_1),
                                 \mathrm{Sub}(\mathrm{X}_1, \mathrm{X}_0))$
       from the LHS of the equation, where
       $\mathrm{Sub} : \mathrm{E} \to \mathrm{E} \to \mathrm{E}$ and
       $\mathrm{X} : \mathbb{N} \to \mathrm{E}$ are constructors of an inductive type $\mathrm{E}$
       describing the syntax.
       This step is called \emph{reification} (also called metaification or quotation).
 \item We normalize $e$ to a formal sum $0 X_0 + 0 X_1$ and check that all its coefficients are
       zero. This decision procedure is implemented and performed inside the proof assistant, and
       its validity is justified by a correctness lemma.
\end{enumerate}
This process (detailed in \sect\ref{sec:large-scale-reflection} and \ref{sec:reification}) applies
to any equation over an Abelian group, and this proof scheme can be adapted to other mathematical
structures, e.g., commutative rings~\cite{DBLP:conf/tacs/Boutin97, DBLP:conf/tphol/GregoireM05},
fields~\cite{coqrefman:ring}, lattices~\cite{DBLP:conf/sfp/JamesH09}, and Kleene
algebras~\cite{DBLP:journals/corr/abs-1105-4537}.

Unfortunately, existing implementations of this proof methodology are known not to cooperate with
\emph{packed classes} very well~\cite{Coq:issue11998, MathComp:issue401}.
The packed classes discipline~\cite{DBLP:conf/tphol/GarillotGMR09} is a methodology to define
mathematical structures in dependent type theory, which allows for the sharing of vocabulary
(definitions and lemmas) across the inheritance hierarchy of structures as well as multiple
inheritance (\sect\ref{sec:canonical-structures} and \ref{sec:packed-classes}).
This methodology is used in the \MC{} library~\cite{mathcomp} for \Coq{} extensively, to provide
more than 70 mathematical structures such as finite groups, rings, fields, as well as their
homomorphisms. 

The source of the incompatibility between proof by large-scale reflection and packed classes is
twofold.
Firstly, packed classes require the proof tools (e.g., the rewriting tactic) to compare overloaded
operators (e.g., the multiplication of rings) modulo conversion to enable the sharing of
vocabulary. This conversion is another instance of computational reflection, so-called
\emph{small-scale reflection}. 
In the case of \MC, the mechanism achieving such term comparison is the \emph{keyed matching}
discipline~\cite{DBLP:conf/itp/GonthierT12} implemented as a part of the \ssr{}
plugin~\cite{coqrefman:ssreflect}.
Secondly, in most of the existing tactics based on large-scale reflection, their reification
procedures recognize operators purely syntactically and do not take conversion into account.
We propose a reification scheme based on keyed matching to address this shortcoming
(\sect\ref{sec:overloaded-reflexive-tactics}).

Another issue is that extending the above reflection scheme to support homomorphisms, whose domain
and codomain types may differ, requires a more involved data type describing the syntax, another
decision procedure, and correctness proof.
In this paper, instead of redefining the syntax and the decision procedure, we propose a new
reflection scheme consisting of two reflection steps.
The first step, which we call \emph{preprocessing}, pushes down homomorphisms in the input terms to
leaves using the structure preservation laws, e.g., $f(x + y) = f(x) + f(y)$. Although preprocessing
requires a heterogeneous syntax that can express a term that has subterms of different types, it
remains quite simple since we do not have to replace variables with numbers in preprocessing as in
$\mathrm{X}_n$ above. 
In the second step, we apply the reflexive decision procedure that uses a homogeneous syntax, as
explained at the beginning of this section.
Since these two kinds of reified terms mostly follow the same syntactic structure, it is possible to
implement a reification procedure that produces both reified terms simultaneously.
Moreover, the preprocessing step allows us to adapt an existing reflexive tactic to operators not
directly supported by its syntax, e.g., opposite which can be expressed as a combination of zero and
subtraction, without modifying the existing syntax, procedures, and correctness proofs
(\sect\ref{sec:extension}).

As an application of our methodology, we adapt the \coq{ring} and \coq{field}
tactics~\cite{DBLP:conf/tphol/GregoireM05, coqrefman:ring} of \Coq{} to the commutative ring and
field structures of \MC, with support for homomorphisms and some operators that cannot be directly
described by the provided syntax (\sect\ref{sec:ring-field}).
Furthermore, we apply our tactics to the formal proof of Ap\'ery's theorem (the irrationality of
$\zeta(3)$ where $\zeta$ is the Riemann zeta function)~\cite{Apery-1979-IDD,
vanderPoorten-1979-PEM} by Chyzak, Mahboubi, and Sibut-Pinote~\cite{apery,DBLP:conf/itp/ChyzakMST14,
DBLP:journals/lmcs/MahboubiS21}, to bring more proof automation (\sect\ref{sec:apery}).
For this purpose, we also reimplemented their technique~%
\cite[\sect{4.3}]{DBLP:conf/itp/ChyzakMST14}\cite[\sect{2.4}]{DBLP:journals/lmcs/MahboubiS21}
to automatically prove proof obligations generated by the \coq{field} tactic using the \coq{lia}
(linear integer arithmetic) tactic~\cite{DBLP:conf/types/Besson06, coqrefman:micromega} of \Coq.
This reimplementation is done based on the approach of Gonthier et
al.~\cite{DBLP:journals/jfp/GonthierZND13} to use canonical structures
(\sect\ref{sec:canonical-structures}) for proof automation, is extensible by declaring canonical
structure instances, and supports a broader range of problems (\sect\ref{sec:non-nullity}).

Our reification procedures are written in \CoqElpi~\cite{coq-elpi} (\sect\ref{sec:reification}).
\Elpi~\cite{DBLP:conf/lpar/DunchevGCT15,elpi} is a dialect of
$\lambda$Prolog~\cite{DBLP:books/daglib/0036008}, a higher-order logic programming language.
The \CoqElpi{} plugin lets us write \Coq{} commands and tactics in \Elpi, and provides a
higher-order abstract syntax (HOAS)~\cite{DBLP:conf/pldi/PfenningE88} embedding of \Coq{} terms in
\Elpi, to manipulate syntax trees with binders in a comfortable way.


\section{Background}
\label{sec:background}

This preliminary section briefly reviews the main ingredients of this paper, namely, canonical
structures (\sect\ref{sec:canonical-structures}), the hierarchy of mathematical structures in \MC{}
(\sect\ref{sec:packed-classes}), large-scale reflection (\sect\ref{sec:large-scale-reflection}), and
reification in \CoqElpi{} (\sect\ref{sec:reification}).

\subsection{Canonical structures}
\label{sec:canonical-structures}

Canonical structures~\cite{DBLP:conf/itp/MahboubiT13, DBLP:phd/hal/Saibi99, coqrefman:canonical}
make it possible to implement ad-hoc inference mechanisms in \Coq{} by giving a particular form of
hints~\cite{DBLP:conf/tphol/AspertiRCT09} to the unification
engine~\cite{DBLP:journals/jfp/ZilianiS17}.
An interface to trigger such an inference is expressed as a record. For example, a record type
declaration
\begin{coqcode}
Structure eqType := { eq_sort : Type; eq_op : eq_sort -> eq_sort -> bool }.
\end{coqcode}
represents a type (\coq{eq_sort}) equipped with a comparison function (\coq{eq_op}).
At the same time, it is an interface to relate a type to its canonical comparison function.
\coq{Structure} is just a synonym of \coq{Record}, but we reserve the former for interfaces for
canonical structure resolution.
A hint can be given as a record instance. For example, an instance
\begin{coqcode}
Canonical nat_eqType : eqType := {| eq_sort := nat; eq_op := eqn |}.
\end{coqcode}
allows us to type check \coq{(@eq_op _ 0
numbers of type \coq{nat} and \coq{eqn} is the comparison function of type
\coq{(nat -> nat -> bool)}.
Since \coq{eq_op} has type \coq{(forall e : eqType, eq_sort e -> ...)}, applying
\coq{eq_op} to \coq{0
check, where \coq{?e} is a unification variable of type \coq{eqType}.
For a \coq{Canonical} declaration, the system synthesizes a unification hint between the projections
(\coq{eq_sort} and \coq{eq_op}) and the head symbols of the fields (\coq{nat} and \coq{eqn}),
respectively. Therefore, the above equation is solved by instantiating \coq{?e} with
\coq{nat_eqType}.

Additionally, declaring the \coq{eq_sort} projection as an implicit
coercion~\cite{DBLP:conf/popl/Saibi97, DBLP:phd/hal/Saibi99, coqrefman:coercions} allows us to use
\coq{(T : eqType)} in the context that expects a term of type \coq{Type}, so that one may write
\coq{(x : T)} rather than \coq{(x : eq_sort T)}.
\begin{coqcode}
Coercion eq_sort : eqType >-> Sortclass.
\end{coqcode}

\subsection{The hierarchy of mathematical structures in \MC}
\label{sec:packed-classes}

\begin{figure}[t]
 \centering
 \begin{tikzpicture}[structure/.style={draw, rounded corners=1mm}, x=8mm, y=7mm]
  \node[structure] (eqType)          at (0,   0) {\coq{eqType}};
  \node[structure] (zmodType)        at (3,   0) {\coq{zmodType}};
  \node[structure] (ringType)        at (6,   0) {\coq{ringType}};
  \node[structure] (comRingType)     at (8,   1) {\coq{comRingType}};
  \node[structure] (unitRingType)    at (8,  -1) {\coq{unitRingType}};
  \node[structure] (comUnitRingType) at (10,  0) {\coq{comUnitRingType}};
  \node[structure] (fieldType)       at (12,  1) {\coq{fieldType}};
  \node[structure] (numDomainType)   at (12, -1) {\coq{numDomainType}};
  \node[structure] (numFieldType)    at (14,  0) {\coq{numFieldType}};
  \draw[->] (eqType)          -- (zmodType);
  \draw[->] (zmodType)        -- (ringType);
  \draw[->] (ringType)        -- (comRingType);
  \draw[->] (ringType)        -- (unitRingType);
  \draw[->] (comRingType)     -- (comUnitRingType);
  \draw[->] (unitRingType)    -- (comUnitRingType);
  \draw[->] (comUnitRingType) -- (fieldType);
  \draw[->] (comUnitRingType) -- (numDomainType);
  \draw[->] (fieldType)       -- (numFieldType);
  \draw[->] (numDomainType)   -- (numFieldType);
 \end{tikzpicture}
 \caption{An excerpt of the hierarchy of mathematical structures in \MC, where an arrow from
 \coq{aType} to \coq{bType} means that \coq{bType} inherits from  \coq{aType}, e.g., \coq{ringType}
 inherits from \coq{zmodType}.}
 \label{fig:hierarchy}
\end{figure}
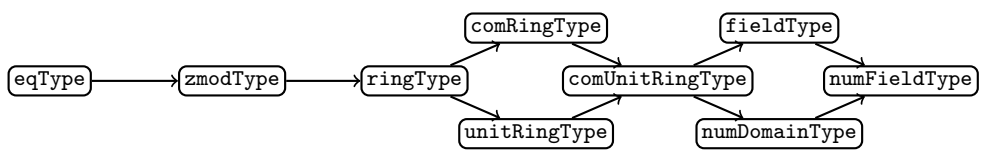

We illustrate a part of the hierarchy of mathematical structures provided by the \MC{} library in
\fig\ref{fig:hierarchy} and summarize its three most basic structures below.
The ones not summarized below are required only in \sect\ref{sec:applications} and explained there.
Each structure is defined as a record bundling a \coq{Type} with operators and axioms as in
\coq{eqType} of \sect\ref{sec:canonical-structures}.
More details on the structures and their operators, including those introduced later, can be found
in \appx\ref{appx:mathcomp-operators} (Table~\ref{table:mathcomp-operators}) and
in~\cite[\chap{2 and 4}]{DBLP:phd/hal/Cohen12}.
\begin{description}
 \item[\coq{(T : eqType)}] is a type whose propositional equality is decidable.
	    The \coq{eqType} record in \sect\ref{sec:canonical-structures} is a simplified version
	    of this structure.
	    For any \coq{x} and \coq{y} of type \coq{T},
	    \coq{(x == y)} ($\coloneqq$ \coq{eq_op x y}) tests if \coq{x}
	    is equal to \coq{y}. Its negation can be expressed as \coq{(x != y)}.
 \item[\coq{(V : zmodType)}] is a $\mathbb{Z}$-module (additive Abelian group).
	    For any \coq{x} and \coq{y} of type \coq{V},
	    \coq{(x + y)} ($\coloneqq$ \coq{GRing.add x y}),
	    \coq{(- x)} ($\coloneqq$ \coq{GRing.opp x}), and
	    \coq{0} ($\coloneqq$ \coq{GRing.zero V})
	    denotes the sum of \coq{x} and \coq{y}, the opposite of \coq{x}, and zero, respectively.
 \item[\coq{(R : ringType)}] is a ring.
	    For any \coq{x} and \coq{y} of type \coq{R},
	    \coq{(x * y)} ($\coloneqq$ \coq{GRing.mul x y}) and
	    \coq{1} ($\coloneqq$ \coq{GRing.one R})
	    denotes the product of \coq{x} and \coq{y}, and one, respectively.
\end{description}
\pagebreak
where ``\coq{E#$_1$#} ($\coloneqq$ \coq{E#$_2$#})'' means that \coq{E#$_1$#} is a
notation~\cite{coqrefman:syntax-extensions} for \coq{E#$_2$#}, and they are syntactically equal.
Each operator above takes a structure instance as its first argument, which is implicit except for
\coq{GRing.zero} and \coq{GRing.one}.

These structures are defined by following packed classes, advocated by Garillot et
al.~\cite{DBLP:conf/tphol/GarillotGMR09} and also detailed in~\cite{DBLP:conf/cade/AffeldtCKMRS20,
DBLP:phd/hal/Garillot11, mathcomp, DBLP:conf/cade/Sakaguchi20}.
For example, the \coq{ringType} structure is defined as follows.
\begin{coqcode}
(* in \textup{Module GRing}: *)
Module Ring.

Record mixin_of (R : zmodType) : Type :=#\label{line:ring:mixin}#
  Mixin { one : R; mul : R -> R -> R; ... (* properties of \textup{one} and \textup{mul} *) }.

Record class_of (R : Type) : Type :=#\label{line:ring:class}#
  Class { base : Zmodule.class_of R; mixin : mixin_of (Zmodule.Pack base) }.

Structure type : Type := Pack { sort : Type; class : class_of sort }.#\label{line:ring:type}#

Definition zmodType (cT : type) : zmodType :=
  @Zmodule.Pack (sort cT) (base (class cT)).

End Ring.
Notation ringType := Ring.type.
\end{coqcode}
The \coq{GRing.Ring} module serves as a namespace qualifying the definitions inside the module,
which are internals to define the \coq{ringType} structure. Each structure has such a module, e.g.,
\coq{GRing.Zmodule} is for \coq{zmodType}.
The structure is divided into three kinds of records: mixin (Line~\ref{line:ring:mixin}), class
(Line~\ref{line:ring:class}), and structure (Line~\ref{line:ring:type}).
%
The mixin record gathers operators and axioms newly introduced by the structure, e.g., the
multiplication, multiplicative identity, and their properties are required to define rings by
extending $\mathbb{Z}$-modules. The class record assembles the mixins of the superclasses. The
structure record is the actual interface of the structure that bundles a \coq{Type} with its class
instance.

\coq{GRing.Ring.zmodType} is an explicit subtyping function that takes a \coq{ringType} and returns
its underlying \coq{zmodType}, which can be made implicit by declaring it as a coercion.
\begin{coqcode}
Coercion Ring.sort : Ring.type >-> Sortclass.
Coercion Ring.zmodType : Ring.type >-> Zmodule.type.
\end{coqcode}
Furthermore, declaring this subtyping function as a canonical instance allows us to write a term
that mixes $\mathbb{Z}$-module and ring operators, e.g., \coq{(0 + 1)}, by solving type equation of
the form \coq{(GRing.Zmodule.sort ?V #$\unify$# GRing.Ring.sort ?R)}.
In general, solving an equation \coq{(GRing.Ring.sort ?R #$\unify$# T)} gives us the ring instance
\coq{?R} of type \coq{T}.
\begin{coqcode}
Canonical Ring.zmodType.
\end{coqcode}

The ring operators are defined by lifting the projections of the mixin record to the structure
record, as follows.
\begin{coqcode}
Definition one (R : ringType) : R := Ring.one (Ring.mixin (Ring.class R)).
Definition mul (R : ringType) : R -> R -> R := Ring.mul (Ring.mixin (Ring.class R)).
\end{coqcode}

Packed classes can also express the hierarchy of morphisms. For example, the \MC{} library provides
the structure \coq|{additive U -> V}| of additive functions ($\mathbb{Z}$-module homomorphisms) from
\coq{U} to \coq{V}.
Its record projection \coq{GRing.Additive.#apply#} 
returns a function of type \coq{(U -> V)} and is used for triggering instance resolution (e.g.,
\sect\ref{sec:extension-homomorphisms}) in the same way as \coq{GRing.Ring.sort} above.
Similarly, there is a structure of ring homomorphisms \coq|{rmorphism R -> S}| which inherits from
additive functions.

\subsection{Large-scale reflection}
\label{sec:large-scale-reflection}

This section demonstrates how to prove $\mathbb{Z}$-module equations by reflection.
Firstly, we define the data type describing the syntax as follows:
\begin{coqcode}
Inductive AGExpr : Type :=
  | AGX : nat -> AGExpr
  | AGO : AGExpr                        (* zero *)
  | AGOpp : AGExpr -> AGExpr            (* opposite *)
  | AGAdd : AGExpr -> AGExpr -> AGExpr. (* addition *)
\end{coqcode}
where \coq{(AGX j)} means \coq{j}-th variable.
This inductive data type allows us to write a \Coq{} function manipulating the syntax. For example,
we can interpret a syntax tree as follows:
\begin{coqcode}
Fixpoint AGeval (V : Type) (zero : V) (opp : V -> V) (add : V -> V -> V)
                (varmap : list V) (e : AGExpr) : V :=
  match e with
    | AGX j => nth zero varmap j
    | AGO => zero
    | AGOpp e1 => opp (AGeval varmap e1)
    | AGAdd e1 e2 => add (AGeval varmap e1) (AGeval varmap e2)
  end.
\end{coqcode}
where the first four arguments are the carrier type and operators of a $\mathbb{Z}$-module, and
\coq{varmap} is a list whose \coq{j}-th item gives the interpretation of the \coq{j}-th variable.
Such an object representing variable assignments is called a \emph{variable map}.

Similarly, we can define a function \coq{AGnorm} of type \coq{(AGExpr -> list int)} that normalizes
a syntax tree to a list of integers representing a formal sum, e.g., $[1; -2]$ represents
$X_0 - 2 X_1$, where \coq{int} is the type of integers defined in \MC.
Their correctness specialized for the case that \coq{V} is \coq{int} can be stated as follows.
\begin{coqcode}
Lemma int_correct (varmap : list int) (e1 e2 : AGExpr) :
  (* if all the coefficients of the normal form of \textup{e1 - e2} is equal to 0, *)
  all (fun i => i == zeroz) (AGnorm (AGAdd e1 (AGOpp e2))) = true ->
  (* \textup{e1} and \textup{e2} evaluated to integers by \textup{AGeval} are equal. *)
  AGeval zeroz oppz addz varmap e1 = AGeval zeroz oppz addz varmap e2.
\end{coqcode}
where \coq{zeroz}, \coq{oppz}, and \coq{addz} are $\mathbb{Z}$-module operators for \coq{int}.

Suppose we want to prove a goal \coq{((x + (- y)) + x = (- y) + (x + x))} for some \coq{(x y : int)}
where \coq{+} and \coq{-} here mean \coq{addz} and \coq{oppz}, respectively.
Thanks to the above reflection lemma, the proof can be done by the following proof term.
\begin{coqcode}
let e1 := AGAdd (AGAdd (AGX 0) (AGOpp (AGX 1))) (AGX 0) in
let e2 := AGAdd (AGOpp (AGX 1)) (AGAdd (AGX 0) (AGX 0)) in
@int_correct [:: x; y] e1 e2 erefl.
\end{coqcode}
Here we used computational reflection twice.
Firstly, \coq{e1} and \coq{e2}, the reified terms representing the LHS and RHS of the goal,
interpreted by \coq{(AGeval ... [:: x; y])} are convertible to the LHS and RHS, respectively.
This conversion is triggered by applying the proof term \coq{(@int_correct ...)} to the goal.
Secondly, the nullity conditions \coq{(all ... = true)} required by the reflection lemma
\coq{int_correct} is checked by reducing its LHS to \coq{true}. This conversion is triggered by
checking if reflexivity (\coq{erefl}) is acceptable as the last argument of \coq{int_correct}.
In the former case, unfolding too many constants may lead to performance issues, and conversion
should be performed carefully. 
In the latter case, we can simply reduce the LHS to \coq{true}, and this is the case where optimized
reduction procedures such as \coq{vm_compute}~\cite{DBLP:conf/icfp/GregoireL02} and
\coq{native_compute}~\cite{DBLP:conf/cpp/BoespflugDG11} can be useful.

\subsection{Implementing reification in \CoqElpi}
\label{sec:reification}

To turn the above method into an automated proof tool, the reified terms and the variable map must
automatically be obtained from the goal.
Since we cannot pattern match on the operators such as \coq{oppz} in the object level, this
reification has to be done in the meta level.

In this section, we implement reification in \CoqElpi. An example of \Elpi{} program follows.
\begin{elpicode}
pred mem o:list term, o:term, o:term. #\label{line:mem:type}#
mem [X|_]  X {{ O }}      :- !. #\label{line:mem:head}#
mem [_|XS] X {{ S lp:N }} :- !, mem XS X N. #\label{line:mem:tail}#
\end{elpicode}
In this code, we define a \emph{predicate} \elpi{mem}.
Line~\ref{line:mem:type} is the type signature of \elpi{mem}, meaning that \elpi{mem} has three
arguments of type \elpi{(list term)}, \elpi{term}, and \elpi{term}, respectively, where \elpi{term}
is the type of \Coq{} terms.
Line~\ref{line:mem:head} and \ref{line:mem:tail} are two \emph{rules} that define the meaning of
\elpi{mem}. Capital identifiers such as \elpi{X}, \elpi{XS}, and \elpi{N} are unification variables.
The syntax \elpi{[X|XS]} is a cons cell of lists whose head and tail are \elpi{X} and \elpi{XS},
respectively. The syntaxes \elpi|{{ ... }}| and \coq{lp:} are the quotation from \Elpi{} to \Coq{}
and the antiquotation from \Coq{} to \Elpi, respectively. Therefore, these two rules are equivalent
to the following:
\begin{elpicode}
mem [X|_]  X (global (indc «O»))          :- !.
mem [_|XS] X (app [global (indc «S»), N]) :- !, mem XS X N.
\end{elpicode}
where \elpi{app} of type \elpi{(list term -> term)} is a constructor of \elpi{term} meaning an
$n$-ary function application of \Coq, and \elpi{(global (indc _))} means a constructor of \Coq.

Actually, the proposition \elpi{(mem XS X N)} asserts that the \elpi{(N + 1)}-th element of
\elpi{XS} is \elpi{X}, where \elpi{N} is a \Coq{} term of type \coq{nat}. Let us consider an example
\elpi{(mem [Y, Z] Z M)}, where \elpi{Y} and \elpi{Z} are distinct \Coq{} terms and \elpi{M} remains
unknown.
The LHS of the first rule requires that the head of \elpi{XS} is \elpi{X}, but this does not apply
to our example. Thus, it attempts matching with the second rule by solving equations
\elpi{[_|XS] = [Y, Z]}, \elpi{X = Z}, and \elpi|{{ S lp:N }} = M|. Then we get \elpi{XS = [Z]} from
the first equation, and proceed to execute its RHS \elpi{(!, mem [Z] Z N)}, which is the conjunction
of the cut (\elpi{!}) operator and \elpi{(mem [Z] Z N)}.
The cut operator prevents backtracking, i.e., trying other rules of \elpi{mem} when the later items
of the conjunction fails.
Since \elpi{(mem [Z] Z N)} matches with the first rule, \elpi{N} is instantiated with
\elpi|{{ O }}|.
In the end, our example \elpi{(mem [Y, Z] Z M)} succeeds with \coq{(S O)} substituted to the
variable \elpi{M}. Indeed, \elpi{Z} is the second element of \elpi{[Y, Z]}.

If the first argument \elpi{XS} is an open-ended list \elpi{[X0, ..., XN | XS']} where \elpi{XS'}
remains unknown, and the given item \elpi{X} is none of the known elements, \elpi{(mem XS X _)}
instantiates \elpi{XS'} with \elpi{[X | _]} and \elpi{X} becomes the \elpi{(N + 2)}-th element of
\elpi{XS}.

We implement reification as a predicate \elpi{quote}, such that \elpi{(quote In Out VarMap)} reifies
\elpi{In} of type \coq{int} to \elpi{Out} of type \coq{AGExpr} and produces an open-ended variable
map \elpi{VarMap}:
\begin{elpicode}
pred quote i:term, o:term, o:list term.
quote {{ zeroz }} {{ AGO }} _ :- !.
quote {{ oppz lp:In1 }} {{ AGOpp lp:Out1 }} VarMap :- !,
  quote In1 Out1 VarMap.
quote {{ addz lp:In1 lp:In2 }} {{ AGAdd lp:Out1 lp:Out2 }} VarMap :- !,
  quote In1 Out1 VarMap, quote In2 Out2 VarMap.
quote In {{ AGX lp:N }} VarMap :- !, mem VarMap In N.
\end{elpicode}
where \texttt{i:} and \texttt{o:} stand for input and output, respectively. Marking an argument as
input avoids instantiation of that argument.
The first three rules of \elpi{quote} are just simple syntactic translation rules for the operators.
If the input does not match with any of those, it should be treated as a variable by the last rule,
which is implemented using the \elpi{mem} predicate above.


\section{Large-scale reflection for packed classes}
\label{sec:overloaded-reflexive-tactics}

Thanks to the techniques reviewed in \sect\ref{sec:large-scale-reflection} and
\ref{sec:reification}, we can implement a tactic \coq{int_zmodule} for solving any integer equation
that holds for any $\mathbb{Z}$-module.
However, its generalization \coq{poly_zmodule} to arbitrary $\mathbb{Z}$-modules, declared as
instances of the \coq{zmodType} structure, is actually not trivial.
First, we describe a naive implementation that fails and analyze the source of the failure in
\sect\ref{sec:syntactic-reification}. Then, we propose a solution to this issue based on the keyed
matching discipline~\cite{DBLP:conf/itp/GonthierT12} in \sect\ref{sec:keyed-reification}.

\subsection{Purely syntactic reification does not work for packed classes}
\label{sec:syntactic-reification}

We first generalize the correctness and reflection lemmas \coq{int_correct} to any
$\mathbb{Z}$-module:
\begin{coqcode}
Lemma AG_norm_subst (V : zmodType) (varmap : list V) (e : AGExpr) :
  AGsubst 0 -

Lemma AG_correct (V : zmodType) (varmap : list V) (e1 e2 : AGExpr) :
  all (fun i => i == 0) (AGnorm (AGAdd e1 (AGOpp e2))) = true ->
  AGeval 0 -
\end{coqcode}
where \coq{AG_norm_subst} is the key lemma to prove \coq{AG_correct}, \coq{AGsubst} is the function
to substitute a variable map to a formal sum, and \coq{-
\coq{GRing.opp} and \coq{GRing.add} implicitly applied to \coq{V}, respectively.

To reimplement the \elpi{quote} predicate, we add a new argument \elpi{V} which is the
\coq{zmodType} instance for the type of the input term, and replace operators \coq{zeroz},
\coq{oppz}, and \coq{addz} with \coq{(@GRing.zero V)}, \coq{(@GRing.opp V)}, and
\coq{(@GRing.add V)}, respectively.
\begin{elpicode}
pred quote i:term, i:term, o:term, o:list term.
quote V {{ @GRing.zero lp:V }} {{ AGO }} _ :- !.
quote V {{ @GRing.opp lp:V lp:In1 }} {{ AGOpp lp:Out1 }} VarMap :- !,
  quote V In1 Out1 VarMap.
quote V {{ @GRing.add lp:V lp:In1 lp:In2 }} {{ AGAdd lp:Out1 lp:Out2 }} VarMap :- !,
  quote V In1 Out1 VarMap, quote V In2 Out2 VarMap.
quote _ In {{ AGX lp:N }} VarMap :- !, mem VarMap In N.
\end{elpicode}

However, this \elpi{quote} predicate fails to reify at least one addition operator in the goal
\coq{(forall x : int, x + 1 = 1 + x)}.
Let us take a closer look at it by \coq{Set Printing All}:
\begin{coqcode}
forall x : int,
@eq (GRing.Zmodule.sort int_ZmodType)
  (@GRing.add #\color{red}{\underline{\texttt{int\_ZmodType}}}# x (GRing.one int_Ring))
  (@GRing.add (#\color{red}{\underline{\texttt{GRing.Ring.zmodType int\_Ring}}}#) (GRing.one int_Ring) x)
\end{coqcode}
where \coq{int_ZmodType} and \coq{int_Ring} are the canonical \coq{zmodType} and \coq{ringType}
instances of \coq{int}, respectively.

The root of the issue is that the two occurrences of \coq{GRing.add} take syntactically different
\coq{zmodType} instances as highlighted in red.
The former instance is inferred from the type of \coq{x}, by solving the type equation
\coq{(GRing.Zmodule.sort ?V #$\unify$# int)}.
The latter instance is inferred from the type of \coq{(GRing.one ?R)} where \coq{?R} is eventually
instantiated with \coq{int_Ring}, by solving the type equation
\coq{(GRing.Zmodule.sort ?V #$\unify$# GRing.Ring.sort ?R)} whose solution is
\coq{(?V := GRing.Ring.zmodType ?R)}.
The \elpi{quote} predicate above requires that all the \coq{zmodType} instances occurring as the
first argument of the operators are syntactically equal to each other.
However, the above goal does not respect this restriction.
In the presence of the inheritance mechanism of packed classes, such syntactically different
instances for the same type and structure coexist~\cite[\sect{3.1}]{DBLP:conf/cade/AffeldtCKMRS20}%
\cite[\sect{2.4}]{DBLP:conf/tphol/GarillotGMR09}\cite[\sect{3}]{DBLP:conf/cade/Sakaguchi20}, and canonical
structure resolution may infer them simultaneously.
Nevertheless, definitional equality of those instances is ensured by \emph{forgetful
inheritance}~\cite{DBLP:conf/cade/AffeldtCKMRS20}, that is, the practice of implementing inheritance
\linebreak and subtyping functions by record inclusion and erasure of some record fields, respectively.



\subsection{Reification by small-scale reflection}
\label{sec:keyed-reification}

Reification recognizing operators by conversion or unification rather than purely syntactic matching
would address the above issue.
However, using full unification for term matching, e.g., triggering unification
\coq{(@GRing.opp V ?t' #$\unify$# t)} to check if \coq{t} is the opposite of an unknown term
\coq{?t'}, can make reification too costly.
Thus, we propose a solution that mixes syntactic matching and conversion as in the keyed matching
discipline~\cite{DBLP:conf/itp/GonthierT12}.
The idea of keyed matching is to find a subterm that matches with a pattern \coq{(f t1 ... tn)} by
attempting the matching operation only on subterms of the form \coq{(f t1' ... tn')}.
While the head constant (the key) \coq{f} has to be the same constant, its arguments can be compared
by conversion or unification.

In our case, the keys are the $\mathbb{Z}$-module operators \coq{GRing.zero}, \coq{GRing.opp}, and
\coq{GRing.add}. The \elpi{quote} predicate can be reimplemented as follows:
\begin{elpicode}
pred quote i:term, i:term, o:term, o:list term.
quote V {{ @GRing.zero lp:V' }} {{ AGO }} _ :- coq.unify-eq V V' ok, !.
quote V {{ @GRing.opp lp:V' lp:In1 }} {{ AGOpp lp:Out1 }} VarMap :- #\label{line:quote:oppr}#
  coq.unify-eq V V' ok, !, quote V In1 Out1 VarMap.
quote V {{ @GRing.add lp:V' lp:In1 lp:In2 }} {{ AGAdd lp:Out1 lp:Out2 }} VarMap :-
  coq.unify-eq V V' ok, !, quote V In1 Out1 VarMap, quote V In2 Out2 VarMap.
quote _ In {{ AGX lp:N }} VarMap :- !, mem VarMap In N.
\end{elpicode}
where \elpi{(coq.unify-eq V V' ok)} asserts that \coq{V} unifies with \coq{V'}. Since the first
argument \coq{V} and the input term do not have any unification variable under normal use of
\coq{quote}, this unification problem falls in a conversion problem that is generally easier and
less costly to solve than unification.
For example, the second rule of \elpi{quote} (Line~\ref{line:quote:oppr}) does not require \coq{V'}
in the input term \coq{(@GRing.opp V' In1)} to be syntactically equal to the first argument \coq{V},
but it compares \coq{V'} with \coq{V} by conversion after syntactic matching of the opposite
operator \coq{GRing.opp}.
Since this conversion is a part of term matching, the cut operator to prevent backtracking comes
after conversion.

The \coq{zmodType} instance to use as the first argument of \elpi{quote} can be obtained by
canonical structure resolution. This inference is implemented as follows.
\begin{elpicode}
pred solve i:goal, o:list sealed-goal.
solve (goal _ _ {{ @eq lp:Ty lp:T1 lp:T2 }} _ _ as G) GS :- #\label{line:solve:matcheq}#
  std.assert-ok! (coq.unify-eq {{ GRing.Zmodule.sort lp:V }} Ty) #\label{line:solve:matchty}#
                 "Cannot find a declared Z-module", !,
  quote V T1 ZE1 VarMap, !, quote V T2 ZE2 VarMap, !, #\label{line:solve:quote}#
  ...
\end{elpicode}
The \elpi{solve} predicate is the entry point of a tactic in \CoqElpi.
The above rule matches the goal proposition with a pattern \coq{(T1 = T2)} where \elpi{T1} and
\elpi{T2} have type \elpi{Ty} (Line~\ref{line:solve:matcheq}), triggers unification
\coq{(GRing.Zmodule.sort V #$\unify$# Ty)} to find the canonical \coq{zmodType} instance \elpi{V} of
\elpi{Ty} (Line~\ref{line:solve:matchty}), and then reifies \elpi{T1} to \elpi{ZE1} and \elpi{T2} to
\elpi{ZE2} using \elpi{V} obtained in the second step (Line~\ref{line:solve:quote}).
Note that if unification by \elpi{coq.unify-eq} fails, its third argument of type \elpi{diagnostic}
carries the error message. The \elpi{std.assert-ok!} predicate of Line~\ref{line:solve:matchty}
asserts that unification given as the first argument succeeds, but if it fails, it prints the
carried error message with the string given as the second argument.

\section{Extending the syntax with homomorphisms and more operators}
\label{sec:extension}

In this section, we implement a new tactic \coq{morph_zmodule}, that extends the syntax supported by
\coq{poly_zmodule} with $\mathbb{Z}$-module homomorphisms (\sect\ref{sec:extension-homomorphisms})
and subtraction (\sect\ref{sec:extension-operators}) which is not directly supported the syntax
\coq{AGExpr}.
These extensions are achieved by adding another layer of reflection which we call
\emph{preprocessing}.
This twofold reflection scheme allows us to reuse the syntax \coq{AGExpr}, interpretation and
normalization procedures \coq{AGeval} and \coq{AGnorm}, and the reflection lemma \coq{AG_correct}
presented in \sect\ref{sec:background} and \ref{sec:overloaded-reflexive-tactics} as is.

\subsection{Homomorphisms}
\label{sec:extension-homomorphisms}

Firstly, we define another inductive type describing the syntax involving homomorphisms.
\begin{coqcode}
Implicit Types (U V : zmodType).

Inductive MExpr : zmodType -> Type := #\label{line:MExpr}#
  | MX V : V -> MExpr V #\label{line:MX}#
  | MO V : MExpr V
  | MOpp V : MExpr V -> MExpr V
  | MAdd V : MExpr V -> MExpr V -> MExpr V
  | MMorph U V : {additive U -> V} -> MExpr U -> MExpr V. #\label{line:MMorph}#
\end{coqcode}
The main difference of this type compared with \coq{AGExpr} is that:
\coq{MExpr} (Line~\ref{line:MExpr}) is parameterized by a \coq{zmodType} instance \coq{V},
the constructor \coq{MX} (Line~\ref{line:MX}) representing a variable takes a term of type \coq{V}
instead of an index of type \coq{nat}, and
the constructor \coq{MMorph} (Line~\ref{line:MMorph}), representing a homomorphism application,
allows for changing the parameter \coq{V}.
Therefore, one shall interpret a reified term of this type without a variable map provided.
\begin{coqcode}
Fixpoint Meval V (e : MExpr V) : V :=
  match e with
    | MX _ x => x
    | MO _ => 0
    | MOpp _ e1 => - Meval e1
    | MAdd _ e1 e2 => Meval e1 + Meval e2
    | MMorph _ _ f e1 => f (Meval e1)
  end.
\end{coqcode}

The normalization procedure we need for \coq{MExpr} is just pushing down homomorphisms
appearing as the \coq{MMorph} constructor to the leaves of the syntax tree:
\begin{coqcode}
Fixpoint Mnorm U V (f : {additive U -> V}) (e : MExpr U) : V :=
  match e in MExpr U return {additive U -> V} -> V with
    | MX _ x => fun f => f x #\label{line:Mnorm:MX}#
    | MO _ => fun _ => 0
    | MOpp _ e1 => fun f => - Mnorm f e1
    | MAdd _ e1 e2 => fun f => Mnorm f e1 + Mnorm f e2
    | MMorph _ _ g e1 => fun f => Mnorm [additive of f \o g] e1 #\label{line:Mnorm:MMorph}#
  end f.
\end{coqcode}
where the third argument \coq|(f : {additive U -> V})| accumulates homomorphisms applied to \coq{e}.
Therefore, the case for \coq{(e := MMorph _ _ g e1)} (Line~\ref{line:Mnorm:MMorph}) constructs a
homomorphism \coq{[additive of f \o g]} that is the function composition of \coq{f} and \coq{g}, and
passes it to the recursive call for normalizing \coq{e1}.
On the other hand, the case for \coq{(e := MX _ x)} (Line~\ref{line:Mnorm:MX}) applies \coq{f} to
the variable \coq{x}.
Since dependent pattern matching on \coq{(e : MExpr U)} forces instantiation of \coq{U} in type
checking of each clause, defining \coq{Mnorm} that type checks requires the so-called \emph{convoy
pattern}~\cite{DBLP:books/daglib/0035083} to propagate this instantiation to the type of \coq{f}.
\pagebreak

Thanks to the structure preservation laws of homomorphisms, a result of normalization
\coq{(Mnorm f e)} should be equal to \coq{f} applied to \coq{(Meval e)}. That is to say, the
following correctness lemma holds:
\begin{coqcode}
Lemma M_correct V (e : MExpr V) : Meval e = Mnorm [additive of idfun] e.
\end{coqcode}
where \coq{[additive of idfun]} is the identity homomorphism.

The reification procedure for the \coq{morph_zmodule} tactic should take an input term \coq{In} of
type \coq{(V : zmodType)}, and obtain a variable map \coq{varmap} and two reified terms \coq{OutM}
and \coq{Out} of types \coq{(MExpr V)} and \coq{AGExpr}, respectively.
For any such \Coq{} terms, the following chain of equations should hold to justify the completeness
of the tactic:
\begin{align*}
 \coq{In}
 &\conveq \coq{Meval OutM}                      & (\text{a meta property}) \\
 &=       \coq{Mnorm [additive of idfun] OutM}  & (\text{Lemma \coq{M_correct}}) \\
 &\conveq \coq{AGeval ... varmap Out}           & (\text{a meta property}) \\
 &=       \coq{AGsubst ... varmap (AGnorm Out)} & (\text{Lemma \coq{AG_norm_subst}})
\end{align*}
where $\conveq$ and $=$ respectively mean definitional equality and propositional equality.
Although these meta properties of reification cannot be proved inside \Coq, the kernel of \Coq{}
will check them for every invocation of the tactic, as explained in
\sect\ref{sec:large-scale-reflection}.

Considering the above requirements, reification can be reimplemented as follows.
\begin{elpicode}
pred quote i:term, i:(term -> term), i:term, o:term, o:term, o:list term.
quote V _ {{ @GRing.zero lp:V' }} {{ MO lp:V }} {{ AGO }} _ :-
  coq.unify-eq V V' ok, !.
quote V F {{ @GRing.opp lp:V' lp:In1 }}
      {{ @MOpp lp:V lp:OutM1 }} {{ AGOpp lp:Out1 }} VarMap :-
  coq.unify-eq V V' ok, !, quote V F In1 OutM1 Out1 VarMap.
quote V F {{ @GRing.add lp:V' lp:In1 lp:In2 }}
      {{ @MAdd lp:V lp:OutM1 lp:OutM2 }} {{ AGAdd lp:Out1 lp:Out2 }} VarMap :-
  coq.unify-eq V V' ok, !,
  quote V F In1 OutM1 Out1 VarMap, quote V F In2 OutM2 Out2 VarMap.
quote V F In {{ @MMorph lp:U lp:V lp:G lp:OutM }} Out VarMap :-  #\label{line:quote:morph}#
  coq.unify-eq {{ @GRing.Additive.apply lp:U lp:V lp:Ph lp:G lp:In1 }} In ok, !,
  quote U (x\ F {{ @GRing.Additive.apply lp:U lp:V lp:Ph lp:G lp:x }})
        In1 OutM Out VarMap.
quote V F In {{ @MX lp:V lp:In }} {{ AGX lp:N }} VarMap :- !,
  mem VarMap (F In) N.
\end{elpicode}
The new \elpi{quote} predicate takes three input arguments: \elpi{V} is a \coq{zmodType} instance,
\elpi{F} is a homomorphism from \elpi{V} to another \coq{zmodType} instance, and \elpi{In} is the
input term of type \elpi{V}.
Then, it produces three output arguments: \elpi{OutM} and \elpi{Out} are the reified terms of types
\coq{(MExpr V)} and \coq{AGExpr}, respectively, and \elpi{VarMap} is the variable map.
The second argument \elpi{F} is required to make recursion of \elpi{quote} work and accumulates
homomorphisms as in the third argument \coq{f} of \coq{Mnorm}. Note that \elpi{F} is represented as
an \Elpi{} function from \elpi{term} to \elpi{term}, which lets us compose functions without leaving
a beta redex in the \Coq{} level.
While the first reified term \elpi{OutM} exactly corresponds to \elpi{In}, the second reified term
\elpi{Out} and the variable map corresponds to \elpi{(F In)}.

The most crucial part of the new \elpi{quote} predicate is its fourth rule
(Line~\ref{line:quote:morph}), which handles the case that the input \elpi{In} is a homomorphism
application.
It first triggers unification \coq{(@GRing.Additive.#apply# U V _ G In1 #$\unify$# In)} to decompose
the input into the homomorphism instance \elpi{G} and its argument \elpi{In1}.
\pagebreak
Then, since \elpi{G} is a homomorphism from \elpi{U} to \elpi{V}, it invokes the recursive call of
\elpi{quote} on \elpi{In1} with \elpi{U} as the first argument (the \coq{zmodType} instance) and the
composition of \elpi{F} and \elpi{G} as the second argument (the homomorphism).
This composition is written as \elpi{(x\ ...)} which means an abstraction $(\lambda x.\, \dots)$.

\subsection{More operators}
\label{sec:extension-operators}

Based on our twofold reflection scheme, we can add support for operators not directly supported by
the syntax \coq{AGExpr}. For example, let \coq{subr} be an opaque subtraction operator of type
\coq{(forall U : zmodType, U -> U -> U)}. By opaque we mean that \coq{subr} does not reduce and thus
we cannot rely on its definitional behavior, but we can reason about it through a lemma:
\begin{coqcode}
subrE : @subr = (fun (U : zmodType) (x y : U) => x + (- y)).
\end{coqcode}

Firstly, we add the following constructor to \coq{MExpr} representing \coq{subr}.
\begin{coqcode}
(* in \textup{Inductive MExpr}: *)
  | MSub V : MExpr V -> MExpr V -> MExpr V
\end{coqcode}

Then, the interpretation and normalization function have to be adapted to the new definition of
\coq{MExpr}, by adding the following cases.
\begin{coqcode}
(* in \textup{Fixpoint Meval}: *)
    | MSub _ e1 e2 => subr (Meval e1) (Meval e2)
\end{coqcode}
\begin{coqcode}
(* in \textup{Fixpoint Mnorm}: *)
    | MSub _ e1 e2 => fun f => Mnorm f e1 + (- Mnorm f e2)
\end{coqcode}

The point is that \coq{Meval} interprets \coq{(MSub V e1 e2)} using \coq{subr}, but \coq{Mnorm}
normalizes it using \coq{GRing.add} and \coq{GRing.opp}.
Proving the correctness lemma \coq{M_correct} based on these new definitions can be done using the
lemma \coq{subrE}. These definitions and the correctness lemma let us replace subexpressions of the
form \coq{(subr e1 e2)} with \coq{(e1' + (- e2'))} by preprocessing, and make it possible to
generate a corresponding reified term of type \coq{AGExpr}.
Therefore, an input term of the form \coq{(subr _ _)} has to be reified to \coq{(@MSub _ _ _)} and
\coq{(AGAdd _ (AGOpp _))}, as follows (see Line~\ref{line:quote:subr:output}).
\begin{elpicode}
quote V F {{ @subr lp:V' lp:In1 lp:In2 }}
      {{ @MSub lp:V lp:OutM1 lp:OutM2 }} {{ AGAdd lp:Out1 (AGOpp lp:Out2) }} #\label{line:quote:subr:output}#
      VarMap :-
  coq.unify-eq V V' ok, !,
  quote V F In1 OutM1 Out1 VarMap, quote V F In2 OutM2 Out2 VarMap.
\end{elpicode}

In fact, even if an operator can be supported by relying on its definitional behavior, our
methodology is sometimes performance-wise better than doing so.
For example, \coq{n
generic embeddings of \coq{(n : nat)} and \coq{(n : int)} to a ring, respectively, where
\coq{(x *+ n)} ($\coloneqq$ \coq{GRing.natmul x n}) and
\coq{(x *~ n)} ($\coloneqq$ \coq{intmul x n})
are \coq{n} times addition of \coq{(x : V)} defined for any \coq{(V : zmodType)}.
For any \coq{n} of type \coq{nat}, \coq{n
latter unfolds to the former, where \coq{Posz} is a constructor of \coq{int} that embeds \coq{nat}
to \coq{int}.
Therefore, if a reflexive tactic supports \coq{n
reifying it in the same way as \coq{(Posz n)
However, it may lead to performance issues by triggering conversions such as:
\begin{coqcode}
Time Check erefl : (Posz 6 * 6)
\end{coqcode}
where \coq{:> rat} means that the LHS and RHS are rational numbers of type \coq{rat}.\footnote{Note
that this particular performance issue reproduces only with \MC{} 1.12.0 or earlier.}

\pagebreak

The source of this inefficiency is that conversion unfolds too many constants.
Computations involving rational numbers are particularly inefficient because \coq{rat} is defined as
a dependent pair of the numerator and denominator that are
coprime~\cite[\sect{4.4.2}]{DBLP:phd/hal/Cohen12} and every \coq{rat} operator performs GCD
calculation to ensure the canonicity of representations.
In our reflection scheme, conversion between \coq{GRing.natmul} and \coq{intmul} can be hidden in
preprocessing. It makes conversion performing only a small number of unfolding and thus more
efficient.

\section{Applications: \coq{ring}, \coq{field}, and the irrationality of $\zeta(3)$}
\label{sec:applications}

As an application of the methodology presented in \sect\ref{sec:overloaded-reflexive-tactics} and
\ref{sec:extension}, we briefly report our effort to adapt the \coq{ring} and \coq{field}
tactics~\cite{DBLP:conf/tphol/GregoireM05, coqrefman:ring} of \Coq{} to the commutative ring and
field structures of \MC{} in \sect\ref{sec:ring-field}.
The \coq{field} tactic generates proof obligations describing the non-nullity of the denominators in
the given equation.
Those conditions can often be simplified to equivalent integer disequations and solved by the
\coq{lia} tactic. In \sect\ref{sec:non-nullity}, we implement this simplification based on the
approach of Gonthier et al.~\cite{DBLP:journals/jfp/GonthierZND13} to use canonical structures for
proof automation.
In \sect\ref{sec:apery}, we apply the above proof tools to the formal proof of Ap\'ery's theorem by
Chyzak, Mahboubi, and Sibut-Pinote~\cite{apery,DBLP:conf/itp/ChyzakMST14,
DBLP:journals/lmcs/MahboubiS21} to bring more proof automation.
Our \coq{ring} and \coq{field} tactics for \MC{} are available as a \Coq{} library called
\emph{Algebra Tactics}~\cite{algebra-tactics}.

We summarize the mathematical structures of \MC{} relevant to this section below. Their inheritance
hierarchy is illustrated in \fig\ref{fig:hierarchy}.
\begin{description}
 \item[\coq{(R : comRingType)}] is a commutative ring.
 \item[\coq{(R : unitRingType)}] is a ring structure with computable inverses.
	    For any \coq{x} of type \coq{R}, \coq{x^-1} ($\coloneqq$ \coq{GRing.inv x}) denotes the
 	    multiplicative inverse of \coq{x}, which is equal to \coq{x} itself if \coq{x} is not a
 	    unit, i.e., has no multiplicative inverse.
 \item[\coq{(R : comUnitRingType)}] is a commutative ring with computable inverses.
 \item[\coq{(F : fieldType)}] is a field.
 \item[\coq{(R : numDomainType)}] is a partially ordered integral domain.
 \item[\coq{(F : numFieldType)}] is a partially ordered field.
\end{description}

\subsection{The \coq{ring} and \coq{field} tactics}
\label{sec:ring-field}

The \coq{ring} and \coq{field} tactics~\cite{DBLP:conf/tphol/GregoireM05, coqrefman:ring} of \Coq{}
respectively solve polynomial and rational equations by computational reflection. Their reflexive
decision procedures are based on normalization to the \emph{sparse Horner
form}~\cite{DBLP:conf/tphol/GregoireM05}, a multivariate, computationally efficient version of the
Horner normal form of polynomials.

The following inductive type describes the syntax supported by the \coq{ring} tactic:
\begin{coqcode}
Inductive PExpr (C : Type) : Type :=
  | PEO : PExpr C                         (* zero:            \textup{GRing.zero} *)
  | PEI : PExpr C                         (* one:             \textup{GRing.one}  *)
  | PEc : C -> PExpr C                    (* constant:        \textup{\_\%:~R}      *)
  | PEX : positive -> PExpr C             (* variable                    *)
  | PEadd : PExpr C -> PExpr C -> PExpr C (* addition:        \textup{GRing.add}  *)
  | PEsub : PExpr C -> PExpr C -> PExpr C (* subtraction:     \textup{\_} \textup{-} \textup{\_}      *)
  | PEmul : PExpr C -> PExpr C -> PExpr C (* multiplication:  \textup{GRing.mul}  *)
  | PEopp : PExpr C -> PExpr C            (* opposite:        \textup{GRing.opp}  *)
  | PEpow : PExpr C -> N -> PExpr C.      (* power:           \textup{GRing.exp}  *)
\end{coqcode}
where \coq{C} is the type of coefficients and fixed to the binary integer type \coq{Z} of the \Coq{}
standard library in our usage. For each constructor, its meaning and the corresponding operator in
\MC{} are indicated in the code comment left.
Note that \coq{(x ^+ n)} ($\coloneqq$ \coq{GRing.exp x n}) is the \coq{n}-th power of \coq{x} with
\coq{(n : nat)}.
There is also \coq{x ^ n} ($\coloneqq$ \coq{exprz x n}) operator, namely, \coq{n}-th power of
\coq{x} with \coq{(n : int)}, which works only for \coq{unitRingType}.

In addition to the above constructs, the \coq{field} tactic supports the following two operators.
\begin{coqcode}
(* in \textup{Inductive FExpr}: *)
  | FEinv : FExpr C -> FExpr C            (* inverse:         \textup{GRing.inv}   *)
  | FEdiv : FExpr C -> FExpr C -> FExpr C (* division:        \textup{\_} \textup{/} \textup{\_}       *)
\end{coqcode}

On top of these syntaxes, we implemented preprocessors to support homomorphisms and more operators
such as \coq{GRing.natmul}, \coq{intmul}, and \coq{exprz}.
Since rings and fields have poorer structures such as $\mathbb{Z}$-modules, subexpressions of these
structures may appear under homomorphism applications.
For example, let \coq{(f : V -> R)} be an additive function whose codomain \coq{R} is a ring, and we
want to perform the following equational reasoning in preprocessing:
\begin{coqcode}
   f (x *~ (n * m))
 = f x * (n * m)
 = f x * (n
\end{coqcode}
This example indicates that ring multiplication may appear in a $\mathbb{Z}$-module subexpression of
a ring expression, and homomorphisms can be pushed down through it.

Therefore, we defined three inductive types describing the syntax: \coq{NExpr} for expressions of
type \coq{nat}, \coq{RExpr} for ring expressions, and \coq{ZMExpr} for $\mathbb{Z}$-module
expressions. The latter two are defined as mutually inductive types.
\coq{RExpr} contains constructors for field operators and is used for both \coq{ring} and
\coq{field} tactics. Since a ring homomorphism can be pushed down through these operators only if
the codomain of the homomorphism is a field, we define normalization functions for \coq{RExpr} and
\coq{ZMExpr} for each of the \coq{ring} and \coq{field} tactics separately.
The definitions of these syntaxes, evaluation and normalization functions, and the correctness
lemmas are available in \appx\ref{appx:preprocessors}.

\subsection{Automating proofs of non-nullity conditions for \coq{field}}
\label{sec:non-nullity}

The \coq{field} tactic can now solve a goal
\begin{coqcode}
((n ^+ 2)
\end{coqcode}
where \coq{F} is a field.
The \coq{field} tactic then generates a proof obligation \coq{(n
non-nullity of the denominator in the equation.
If \coq{F} is a partially ordered field (\coq{numFieldType}), this obligation can be simplified to
\coq{(n != 1 :> int)} because \coq{_
(\coq{numDomainType}) is injective.
The simplified obligation can sometimes be solved by other automated tactics such as
\coq{lia}~\cite{DBLP:conf/types/Besson06, coqrefman:micromega}, which can solve linear goals over
integers.
Note that applying the \coq{lia} tactic to formulae stated using the arithmetic operators of \MC{}
requires another preprocessing, which we reimplemented as another small library called
Mczify~\cite{mczify}.
This combination of the \coq{field} and \coq{lia} tactics is extensively used in the formal proof of
Ap\'ery's theorem~\cite[\sect{4.3}]{DBLP:conf/itp/ChyzakMST14}%
\cite[\sect{2.4}]{DBLP:journals/lmcs/MahboubiS21}.

In this section, we reimplement this simplification based on the approach
of Gonthier et al.~\cite{DBLP:journals/jfp/GonthierZND13}.
Firstly, we define a canonical structure \coq{zifyRing} that relates a ring expression that is an
element of the integer subring (\coq{rval : R}), to the corresponding integer expression
(\coq{zval : int}) such that \coq{(rval = zval
\begin{coqcode}
Section ZifyRing.

Variable R : ringType.

Structure zifyRing :=
  ZifyRing { rval : R; zval : int; zifyRingE : rval = zval
\end{coqcode}

For instance, the \coq{zifyRing} record allows us to relate \coq{(0 : R)} to \coq{(0 : int)} and
\coq{(1 : R)} to \coq{(1 : int)}.
\begin{coqcode}
Canonical zify_zero := @ZifyRing 0 0 (erefl : 0 = 0
Canonical zify_one  := @ZifyRing 1 1 (erefl : 1 = 1
\end{coqcode}
Since the integer subring is closed under opposite, \coq{(- x = (- n)
\coq{(x = n
This implication can be encoded as a canonical instance that takes another instance as an argument,
as follows.
\begin{coqcode}
Lemma zify_opp_subproof (e1 : zifyRing) : - rval e1 = (- zval e1)

Canonical zify_opp (e1 : zifyRing) :=
  @ZifyRing (- rval e1) (- zval e1) (zify_opp_subproof e1).
\end{coqcode}
Similarly, the closure properties under \coq{GRing.add}, \coq{GRing.mul}, and \coq{intmul} can be
implemented as the following instances.
\begin{coqcode}
Canonical zify_add e1 e2 := @ZifyRing (rval e1 + rval e2) (zval e1 + zval e2) ...
Canonical zify_mul e1 e2 := @ZifyRing (rval e1 * rval e2) (zval e1 * zval e2) ...
Canonical zify_mulrz e1 n := @ZifyRing (rval e1 *~ n) (zval e1 *~ n) ...
\end{coqcode}

In general, solving an equation \coq{(rval ?e1 #$\unify$# x)} gives us an integer expression \coq{n}
and its correctness proof of \coq{(x = n
example \coq{(x := 1 + n
Solving the equation \coq{(rval ?e1 #$\unify$# x)} proceeds by instantiating \coq{?e1} with
\coq{(zify_add ?e2 ?e3)} since the head symbol of \coq{x} is \coq{GRing.add}, and then the problem
is divided into two sub-problems \coq{(rval ?e2 #$\unify$# 1)} and
\coq{(rval ?e3 #$\unify$# n
Solving the former sub-problem is done by instantiating \coq{?e2} with \coq{zify_one}, solving the
latter proceeds by instantiating \coq{?e3} with \coq{(zify_mulrz ?e4 2)}, and we get another
sub-problem \coq{(rval ?e4 #$\unify$# n
By repeating this recursive process, we eventually get the canonical solution
\coq{(?e4 := zify_mulrz zify_one n)}.
The \coq{zval} and \coq{zifyRingE} fields of the solution
\coq{(?e1 := zify_add zify_one (zify_mulrz (zify_mulrz zify_one n) 2))} give us the integer
expression and the proof, respectively.

Reducing a ring (dis)equation to an integer (dis)equation is performed by rewriting the ring
equation by the following lemma.
\begin{coqcode}
End ZifyRing.

Lemma zify_eqb (R : numDomainType) (e1 e2 : zifyRing R) :
  (rval e1 == rval e2) = (zval e1 == zval e2).
\end{coqcode}
For example, combining the above lemma and the \coq{lia} tactic allows us to solve the following
goal. We use a small \Ltac~\cite{DBLP:conf/lpar/Delahaye00} script to perform this proof automation
in practice.
\begin{coqcode}
Goal forall n : int, n
Proof. move=> n; rewrite zify_eqb /=; lia. Qed.
\end{coqcode}

\subsection{The irrationality of $\zeta(3)$}
\label{sec:apery}

This section briefly reports the result of applying the proof tools presented in the previous
sections to the formal proof of Ap\'ery's theorem~\cite{apery,DBLP:conf/itp/ChyzakMST14,
DBLP:journals/lmcs/MahboubiS21}.
This proof involves various collections of numbers such as integers, rational numbers \coq{rat},
their real closure \coq{realalg}, algebraic numbers \coq{algC}, and Cauchy sequences.
These types are equipped with ring instances except for Cauchy sequences and also with field
instances except for integers. Therefore, the embedding functions corresponding to their inclusion,
e.g., $\mathbb{Z} \subset \mathbb{Q}$, are ring homomorphisms.
Since the type of integers \coq{int} of \MC{} is defined based on Peano natural numbers \coq{nat},
it is well suited for proofs but prevents us from performing computation involving large integer
constants in a reasonable time.
Therefore, this proof also uses the binary representation of integers \coq{Z} for computation
purposes and defines a function that embeds \coq{Z} to rational numbers \coq{(rat_of_Z : Z -> rat)}.
This embedding function is made opaque to prevent computing in \coq{rat}.

We managed to replace two tactics in this proof with our tools: \coq{rat_field} adapting the
\coq{field} tactic to \MC, and \coq{goal_to_lia} implementing the reduction of
\sect\ref{sec:non-nullity}, both of which are implemented in \Ltac{} and specific to rational
numbers \coq{rat}.
This replacement is done by adding the support for \coq{Z} constants and operators to our
\coq{field} tactic and by making \coq{rat_of_Z} canonically a ring homomorphism.
That is to say, we did not have to implement any treatment specific to this proof to our tools,
since supporting large integer constants is considered to be of general interest.
Moreover, our \coq{ring} and \coq{field} tactics can reason about any ring and field instances and
ring homomorphisms. Thus, they can solve a broader range of subgoals, and some manual labor before
or after invoking them, e.g., tweaking ring homomorphisms, has been handed off to our tools.

\begin{table}[t]
 \centering
 \caption{Performance comparison of the \coq{rat_field} tactic and our \coq{field} tactic in
 \texttt{ops\_for\_b.v} of the formal proof of Ap\'ery's theorem, which proves a recurrence equation
 satisfied by a sequence called $b$~\cite[\sect{4}]{DBLP:conf/itp/ChyzakMST14}.
 The size of a problem is the number of constructors of reified terms of type \coq{FExpr}. Note that
 Problem \#1 has many relatively large coefficients greater than $10^{12}$.
 Therefore, the ratio of time spent for its normalization is higher than those of the other
 problems, and thus its improvement in execution time is relatively minor.}
 \label{table:apery-performance}
 \begin{tabular}{llrrr}
  Lemma              & \# Problem & Size    & \coq{rat_field} & \coq{field} \\
                     &            &         & Time (s)        & Time (s) \\ \hline
  \coq{P_eq_Delta_Q} & 1          &  14,690 & 91.872          & 87.410 \\
  \coq{recAperyB}    & 2          &   8,407 &  4.168          &  1.957 \\
  \coq{recAperyB}    & 3          & 113,657 & 39.033          & 26.680
 \end{tabular}
\end{table}

Our tools not only automate more proofs but also, directly and indirectly, make proofs more concise
and faster to check.
To give some figures, Table~\ref{table:apery-performance} summarizes the performance of the
invocations of \coq{rat_field} and \coq{field} that take more than 1 second in the proof. In those
cases, \coq{field} is consistently faster than \coq{rat_field}.
Moreover, by extensively refactoring proofs using our tools, we could reduce 485 lines of
specifications and proofs out of 5881 lines excluding code for proof automation, and checking the
entire proof became 26\% faster (6 min 52 s) than before (9 min 19 s).

On the other hand, we still see some room for improvement in this refactoring work.
For example, our \coq{field} tactic cannot directly solve an equation that has rational exponents,
e.g., $x^{\frac{3}{2}}$, or variables in exponents~\cite{DBLP:conf/cade/Baanen20}, e.g.,
$x^{n + m} = x^n x^m$.
However, they require reimplementing reflexive decision procedures and are pretty orthogonal to the
present work, except that it might be possible to implement incomplete support for the latter case
in preprocessing.

\section{Conclusion}
\label{sec:conclusion}

We proposed a methodology for building reflexive tactics and their concrete implementations in
\CoqElpi{} that cooperate with algebraic structures (\sect\ref{sec:overloaded-reflexive-tactics})
and their homomorphisms (\sect\ref{sec:extension-homomorphisms}) represented by packed classes.
The issue we solved in \sect\ref{sec:overloaded-reflexive-tactics} is not specific to packed
classes, as the issue solved by forgetful inheritance~\cite{DBLP:conf/cade/AffeldtCKMRS20} also
appears in semi-bundled~\cite[\sect{4.1}]{DBLP:conf/cpp/X20} type
classes~\cite{DBLP:conf/tphol/SozeauO08}.
On the other hand, purely syntactic reification works fine with
unbundled~\cite{DBLP:journals/mscs/SpittersW11} type classes, where operators appear as parameters
of interfaces, as is the case in~\cite{DBLP:conf/cpp/BraibantP11,
DBLP:journals/corr/abs-1105-4537}. However, this approach does not scale up to larger hierarchies,
e.g., as noted in~\cite[\sect{6.1}]{DBLP:journals/corr/abs-1105-4537}.
Reification by small-scale reflection can also be adapted to reification by
parametricity~\cite{DBLP:conf/itp/GrossEC18}, although it does not deal with variable maps and thus
does not fit our purpose.
Such implementation can be done by using the \coq{ssrpattern} tactic in place of the \coq{pattern}
tactic, but it may not preserve the efficiency of reification by parametricity.

We argue that \CoqElpi{} turned out to be a practical tool to implement our methodology, and in
particular, provides features that made our reification procedures concise, although we could
reimplement our tactics with other meta-languages such as OCaml,
\Ltac~\cite{DBLP:conf/lpar/Delahaye00}, \Ltactwo~\cite{Pedrot:Ltac2}, and
\Mtactwo~\cite{DBLP:journals/pacmpl/KaiserZKRD18}.
For example, the cut operator, which is unavailable in \Ltac, offers a pretty intuitive way to
control backtracking, and quotation and antiquotation allow us to embed \Coq{} terms with holes to
our \Elpi{} code in a readable way.
Moreover, our resulting tactics run in reasonable times.

Our twofold reflection scheme and its preprocessing step to support homomorphisms allow us to adapt
an existing reflexive tactic to new operators without either reimplementing the whole tactic or
reifying similar terms twice (\sect\ref{sec:extension-operators}).
However, it does not let users extend an existing preprocessor with new rules as the \coq{ppsimpl}
tactic~\cite{Besson:ppsimpl} does, although \coq{ppsimpl} is not flexible enough to cover our use
cases.
Since \CoqElpi{} provides the abilities to generate inductive data types, \Coq{} constants, and
\Elpi{} rules, we could improve this situation by writing an \Elpi{} program that produces a
reflexive preprocessor and reification rules from their high-level descriptions.
Furthermore, we could integrate such an enhancement to Hierarchy
Builder~\cite{DBLP:conf/fscd/CohenST20} to utilize metadata about the hierarchy of structures in
reification.

As an application of our methodology, we adapted the \coq{ring} and \coq{field} tactics of \Coq{} to
the commutative rings and fields of \MC{} (\sect\ref{sec:ring-field}).
We demonstrated their practicality and scalability by applying them to the formal proof of Ap\'ery's
theorem (\sect\ref{sec:apery}).
Although their reflexive decision procedures are not our contribution, we found room for improvement
on this point.
For example, the \texttt{ring\_exp} tactic~\cite{DBLP:conf/cade/Baanen20} of
\Lean~\cite{DBLP:conf/cade/MouraKADR15} solves ring equations with variables in exponents, which is
one of the cases we wished to solve in \sect\ref{sec:apery}.
The \texttt{ring\_exp} tactic does not directly support homomorphisms, but the \texttt{simp} and
\texttt{norm\_cast} tactics~\cite{DBLP:conf/cade/LewisM20} serve as preprocessors for pushing down
and up homomorphisms as in \sect\ref{sec:extension-homomorphisms} and \ref{sec:non-nullity}.
In contrast to our approach, those \Lean{} tactics can be performed alone, which is an easier way to
achieve modularity of proof tools, but then, each tactic has to traverse the goal.
Also, they are not reflexive and produce proof terms explaining rewriting steps.
While such implementation does not require proving the procedure correct and is regarded as
performance-wise better than reflection in \Lean, it would not scale up to large equations and
expressions that have large normal forms.
On the other hand, implementing efficient tactics using computational reflection requires verified
and efficient procedures involving computation-oriented data structures such as sparse Horner
form~\cite{DBLP:conf/tphol/GregoireM05}.
Cohen and Rouhling~\cite{cohen:hal-01414881} proposed a modular approach to define and reason about
efficient decision procedures using \CoqEAL{} refinement
framework~\cite{DBLP:conf/cpp/CohenDM13, DBLP:conf/itp/DenesMS12}, which is a suitable candidate
method for extensively developing reflexive tactics for mathematical structures of \MC.

\bibliography{bibliography}

\appendix

\section{Descriptions of structures and operators}
\label{appx:mathcomp-operators}

We summarize the mathematical structures and their operators relevant to this paper in
Table~\ref{table:mathcomp-operators}. Besides those, each \coq{#\textit{struct}Type#} ($\coloneqq$
\coq{#\textit{Struct}#.type}) structure is equipped with:
\begin{itemize}
 \item an implicit coercion \coq{#\textit{Struct}#.sort} from \coq{#\textit{Struct}#.type} to types
       (\coq{Sortclass}) that returns the underlying carrier of a structure instance, and
 \item a notation \coq{[#\textit{struct}Type# of T]} that gives the canonical
       \coq{#\textit{struct}Type#} instance of \coq{T} if it exists; otherwise, it does not
       type-check.
\end{itemize}

\section{Reflexive preprocessors for the \coq{ring} and \coq{field} tactics}
\label{appx:preprocessors}

\begin{coqcode}
Implicit Types (V : zmodType) (R : ringType) (F : fieldType).

Inductive NExpr : Type :=
  | NC    of N
  | NX    of nat
  | NAdd  of NExpr & NExpr
  | NSucc of NExpr
  | NMul  of NExpr & NExpr
  | NExp  of NExpr & N.

Fixpoint Neval (e : NExpr) : nat :=
  match e with
    | NC n => nat_of_N_expand n
    | NX x => x
    | NAdd e1 e2 => Neval e1 + Neval e2
    | NSucc e => S (Neval e)
    | NMul e1 e2 => Neval e1 * Neval e2
    | NExp e1 n => Neval e1 ^ nat_of_N_expand n
  end.

Inductive RExpr : ringType -> Type :=
  | RX R : R -> RExpr R
  | R0 R : RExpr R
  | ROpp R : RExpr R -> RExpr R
  | RZOpp : RExpr [ringType of Z] -> RExpr [ringType of Z]
  | RAdd R : RExpr R -> RExpr R -> RExpr R
  | RZAdd : RExpr [ringType of Z] -> RExpr [ringType of Z] ->
            RExpr [ringType of Z]
  | RZSub : RExpr [ringType of Z] -> RExpr [ringType of Z] ->
            RExpr [ringType of Z]
  | RMuln R : RExpr R -> NExpr -> RExpr R
  | RMulz R : RExpr R -> RExpr [ringType of int] -> RExpr R
  | R1 R : RExpr R
  | RMul R : RExpr R -> RExpr R -> RExpr R
  | RZMul : RExpr [ringType of Z] -> RExpr [ringType of Z] ->
            RExpr [ringType of Z]
  | RExpn R : RExpr R -> N -> RExpr R
  | RExpPosz (R : unitRingType) : RExpr R -> N -> RExpr R
  | RExpNegz F : RExpr F -> N -> RExpr F
  | RZExp : RExpr [ringType of Z] -> Z -> RExpr [ringType of Z]
  | RInv F : RExpr F -> RExpr F
  | RMorph R' R : {rmorphism R' -> R} -> RExpr R' -> RExpr R
  | RMorph' V R : {additive V -> R} -> ZMExpr V -> RExpr R
  | RPosz : NExpr -> RExpr [ringType of int]
  | RNegz : NExpr -> RExpr [ringType of int]
  | RZC : Z -> RExpr [ringType of Z]
with ZMExpr : zmodType -> Type :=
  | ZMX V : V -> ZMExpr V
  | ZM0 V : ZMExpr V
  | ZMOpp V : ZMExpr V -> ZMExpr V
  | ZMAdd V : ZMExpr V -> ZMExpr V -> ZMExpr V
  | ZMMuln V : ZMExpr V -> NExpr -> ZMExpr V
  | ZMMulz V : ZMExpr V -> RExpr [ringType of int] -> ZMExpr V
  | ZMMorph V' V : {additive V' -> V} -> ZMExpr V' -> ZMExpr V.

Fixpoint Reval R (e : RExpr R) : R :=
  match e with
    | RX _ x => x
    | R0 _ => 0
    | ROpp _ e1 => - Reval e1
    | RZOpp e1 => Z.opp (Reval e1)
    | RAdd _ e1 e2 => Reval e1 + Reval e2
    | RZAdd e1 e2 => Z.add (Reval e1) (Reval e2)
    | RZSub e1 e2 => Z.sub (Reval e1) (Reval e2)
    | RMuln _ e1 e2 => Reval e1 *+ Neval e2
    | RMulz _ e1 e2 => Reval e1 *~ Reval e2
    | R1 _ => 1
    | RMul _ e1 e2 => Reval e1 * Reval e2
    | RZMul e1 e2 => Z.mul (Reval e1) (Reval e2)
    | RExpn _ e1 n => Reval e1 ^+ nat_of_N_expand n
    | RExpPosz _ e1 n => Reval e1 ^ Posz (nat_of_N_expand n)
    | RExpNegz _ e1 n => Reval e1 ^ Negz (nat_of_N_expand n)
    | RZExp e1 n => Z.pow (Reval e1) n
    | RInv _ e1 => (Reval e1)^-1
    | RMorph _ _ f e1 => f (Reval e1)
    | RMorph' _ _ f e1 => f (ZMeval e1)
    | RPosz e1 => Posz (Neval e1)
    | RNegz e2 => Negz (Neval e2)
    | RZC x => x
  end
with ZMeval V (e : ZMExpr V) : V :=
  match e with
    | ZMX _ x => x
    | ZM0 _ => 0
    | ZMOpp _ e1 => - ZMeval e1
    | ZMAdd _ e1 e2 => ZMeval e1 + ZMeval e2
    | ZMMuln _ e1 e2 => ZMeval e1 *+ Neval e2
    | ZMMulz _ e1 e2 => ZMeval e1 *~ Reval e2
    | ZMMorph _ _ f e1 => f (ZMeval e1)
  end.

Section Rnorm.

Variables (R' : ringType).
Variables (R_of_Z : Z -> R') (R_of_ZE : R_of_Z = (fun n => (int_of_Z n)
Variables (zero : R') (zeroE : zero = 0
Variables (add : R' -> R' -> R') (addE : add = +
Variables (sub : R' -> R' -> R') (subE : sub = (fun x y => x - y)).
Variables (one : R') (oneE : one = 1
Variables (mul : R' -> R' -> R') (mulE : mul = *
Variables (exp : R' -> N -> R') (expE : exp = (fun x n => x ^+ nat_of_N n)).

Fixpoint Nnorm (e : NExpr) : R' :=
  match e with
    | NC N0 => R_of_Z Z0
    | NC (Npos n) => R_of_Z (Zpos n)
    | NX x => x
    | NAdd e1 e2 => add (Nnorm e1) (Nnorm e2)
    | NSucc e1 => add one (Nnorm e1)
    | NMul e1 e2 => mul (Nnorm e1) (Nnorm e2)
    | NExp e1 n => exp (Nnorm e1) n
  end.

Fixpoint Rnorm R (f : {rmorphism R -> R'}) (e : RExpr R) : R' :=
  match e in RExpr R return {rmorphism R -> R'} -> R' with
    | RX _ x => fun f => f x
    | R0 _ => fun => zero
    | ROpp _ e1 => fun f => opp (Rnorm f e1)
    | RZOpp e1 => fun f => opp (Rnorm f e1)
    | RAdd _ e1 e2 => fun f => add (Rnorm f e1) (Rnorm f e2)
    | RZAdd e1 e2 => fun f => add (Rnorm f e1) (Rnorm f e2)
    | RZSub e1 e2 => fun f => sub (Rnorm f e1) (Rnorm f e2)
    | RMuln _ e1 e2 => fun f => mul (Rnorm f e1) (Nnorm e2)
    | RMulz _ e1 e2 => fun f =>
        mul (Rnorm f e1) (Rnorm [rmorphism of intmul 1] e2)
    | R1 _ => fun => one
    | RMul _ e1 e2 => fun f => mul (Rnorm f e1) (Rnorm f e2)
    | RZMul e1 e2 => fun f => mul (Rnorm f e1) (Rnorm f e2)
    | RExpn _ e1 n => fun f => exp (Rnorm f e1) n
    | RExpPosz _ e1 n => fun f => exp (Rnorm f e1) n
    | RExpNegz _ _ _ => fun _ => f (Reval e)
    | RZExp e1 (Z.neg _) => fun f => zero
    | RZExp e1 n => fun f => exp (Rnorm f e1) (Z.to_N n)
    | RInv _ _ => fun _ => f (Reval e)
    | RMorph _ _ g e1 => fun f => Rnorm [rmorphism of f \o g] e1
    | RMorph' _ _ g e1 => fun f => RZMnorm [additive of f \o g] e1
    | RPosz e1 => fun => Nnorm e1
    | RNegz e1 => fun => opp (add one (Nnorm e1))
    | RZC x => fun => R_of_Z x
  end f
with RZMnorm V (f : {additive V -> R'}) (e : ZMExpr V) : R' :=
  match e in ZMExpr V return {additive V -> R'} -> R' with
    | ZMX _ x => fun f => f x
    | ZM0 _ => fun => zero
    | ZMOpp _ e1 => fun f => opp (RZMnorm f e1)
    | ZMAdd _ e1 e2 => fun f => add (RZMnorm f e1) (RZMnorm f e2)
    | ZMMuln _ e1 e2 => fun f => mul (RZMnorm f e1) (Nnorm e2)
    | ZMMulz _ e1 e2 => fun f =>
        mul (RZMnorm f e1) (Rnorm [rmorphism of intmul 1] e2)
    | ZMMorph _ _ g e1 => fun f => RZMnorm [additive of f \o g] e1
  end f.

Lemma Rnorm_correct (e : RExpr R') : Reval e = Rnorm [rmorphism of idfun] e.

End Rnorm.

Section Fnorm.

Variables (F : fieldType).
Variables (F_of_Z : Z -> F) (F_of_ZE : F_of_Z = (fun n => (int_of_Z n)
Variables (zero : F) (zeroE : zero = 0
Variables (add : F -> F -> F) (addE : add = +
Variables (sub : F -> F -> F) (subE : sub = (fun x y => x - y)).
Variables (one : F) (oneE : one = 1
Variables (exp : F -> N -> F) (expE : exp = (fun x n => x ^+ nat_of_N n)).
Variables (inv : F -> F) (invE : inv = GRing.inv).

Notation Nnorm := (Nnorm F_of_Z add one mul exp).

Fixpoint Fnorm R (f : {rmorphism R -> F}) (e : RExpr R) : F :=
  match e in RExpr R return {rmorphism R -> F} -> F with
    | RX _ x => fun f => f x
    | R0 _ => fun => zero
    | ROpp _ e1 => fun f => opp (Fnorm f e1)
    | RZOpp e1 => fun f => opp (Fnorm f e1)
    | RAdd _ e1 e2 => fun f => add (Fnorm f e1) (Fnorm f e2)
    | RZAdd e1 e2 => fun f => add (Fnorm f e1) (Fnorm f e2)
    | RZSub e1 e2 => fun f => sub (Fnorm f e1) (Fnorm f e2)
    | RMuln _ e1 e2 => fun f => mul (Fnorm f e1) (Nnorm e2)
    | RMulz _ e1 e2 => fun f =>
        mul (Fnorm f e1) (Fnorm [rmorphism of intmul 1] e2)
    | R1 _ => fun => one
    | RMul _ e1 e2 => fun f => mul (Fnorm f e1) (Fnorm f e2)
    | RZMul e1 e2 => fun f => mul (Fnorm f e1) (Fnorm f e2)
    | RExpn _ e1 n => fun f => exp (Fnorm f e1) n
    | RExpPosz _ e1 n => fun f => exp (Fnorm f e1) n
    | RExpNegz _ e1 n => fun f => inv (exp (Fnorm f e1) (N.succ n))
    | RZExp e1 (Z.neg _) => fun f => zero
    | RZExp e1 n => fun f => exp (Fnorm f e1) (Z.to_N n)
    | RInv _ e1 => fun f => inv (Fnorm f e1)
    | RMorph _ _ g e1 => fun f => Fnorm [rmorphism of f \o g] e1
    | RMorph' _ _ g e1 => fun f => FZMnorm [additive of f \o g] e1
    | RPosz e1 => fun => Nnorm e1
    | RNegz e1 => fun => opp (add one (Nnorm e1))
    | RZC x => fun => F_of_Z x
  end f
with FZMnorm V (f : {additive V -> F}) (e : ZMExpr V) : F :=
  match e in ZMExpr V return {additive V -> F} -> F with
    | ZMX _ x => fun f => f x
    | ZM0 _ => fun => zero
    | ZMOpp _ e1 => fun f => opp (FZMnorm f e1)
    | ZMAdd _ e1 e2 => fun f => add (FZMnorm f e1) (FZMnorm f e2)
    | ZMMuln _ e1 e2 => fun f => mul (FZMnorm f e1) (Nnorm e2)
    | ZMMulz _ e1 e2 => fun f =>
        mul (FZMnorm f e1) (Fnorm [rmorphism of intmul 1] e2)
    | ZMMorph _ _ g e1 => fun f => FZMnorm [additive of f \o g] e1
  end f.

Lemma Fnorm_correct (e : RExpr F) : Reval e = Fnorm [rmorphism of idfun] e.

End Fnorm.
\end{coqcode}

\begin{landscape}
 \begin{longtable}{r@{\,\,\,}l@{\,\,\,}lm{7cm}}
  \caption{An excerpt of the descriptions of structures and operators in \MC{}.}
  \label{table:mathcomp-operators}\\
  \multicolumn{2}{c}{\Coq{} judgements} & Synonym of & Informal semantics \\
  \hline
  & \coq|#$\vdash$# T : eqType| &
    (RHS) \coq|#$\coloneqq$# Equality.type| &
    \coq|T| is a \coq|Type| whose propositional (Leibniz) equality is decidable. \\
  %
  %
  \coq|T : eqType, x, y : T| & \coq|#$\vdash$# x == y : bool| &
    \coq|#$\coloneqq$# #\color{gray}\texttt{@}#eq_op #\color{gray}\texttt{T}# x y| &
    a Boolean test to decide if \coq|x| is equal to \coq|y|. \\
  \coq|T : eqType, x, y : T| & \coq|#$\vdash$# x != y : bool| &
    \coq|#$\coloneqq$# ~~ (x == y)| &
    a Boolean test to decide if \coq|x| is not equal to \coq|y|. \\
  \hline
  & \coq|#$\vdash$# V : zmodType| &
    (RHS) \coq|#$\coloneqq$# GRing.Zmodule.type| &
    \coq|V| is a $\mathbb{Z}$-module, i.e., an additive Abelian group. \\
  %
  %
  \coq|V : zmodType| & \coq|#$\vdash$# 0 : V| &
    \coq|#$\coloneqq$# GRing.zero V| &
    the zero (additive identity) of \coq|V|. \\
  \coq|V : zmodType, x : V| & \coq|#$\vdash$# - x : V| &
    \coq|#$\coloneqq$# #\color{gray}\texttt{@}#GRing.opp #\color{gray}\texttt{V}# x| &
    the opposite (additive inverse) of \coq|x| in \coq|V|. \\
  \coq|V : zmodType, x, y : V| & \coq|#$\vdash$# x + y : V| &
    \coq|#$\coloneqq$# #\color{gray}\texttt{@}#GRing.add #\color{gray}\texttt{V}# x y| &
    the sum of \coq|x| and \coq|y| in \coq|V|. \\
  \coq|V : zmodType, x, y : V| & \coq|#$\vdash$# x - y : V| &
    \coq|#$\coloneqq$# x + (- y)| &
    the difference of \coq|x| and \coq|y| in \coq|V|. \\
  $\begin{array}{@{}l@{}} \coq|V : zmodType,| \\ \qquad\coq|x : V, n : nat| \end{array}$ &
    \coq|#$\vdash$# x *+ n : V| &
    \coq|#$\coloneqq$# #\color{gray}\texttt{@}#GRing.natmul #\color{gray}\texttt{V}# x n| &
    \coq|n| times \coq|x| with \coq|(n : nat)|. \\
  $\begin{array}{@{}l@{}} \coq|V : zmodType,| \\ \qquad\coq|x : V, n : nat| \end{array}$ &
    \coq|#$\vdash$# x *- n : V| &
    \coq|#$\coloneqq$# - (x *+ n)| &
    the opposite of \coq|(x *+ n)|. \\
  $\begin{array}{@{}l@{}} \coq|V : zmodType,| \\ \qquad\coq|x : V, n : int| \end{array}$ &
    \coq|#$\vdash$# x *~ n : V| &
    \coq|#$\coloneqq$# #\color{gray}\texttt{@}#intmul #\color{gray}\texttt{V}# x n| &
    \coq|n| times \coq|x| with \coq|(n : int)|,
    reduces to either \coq|(x *+ _)| or  \coq|(x *- _)|. \\
  \hline
  \coq|U, V : zmodType| & \coq|#$\vdash$# f : {additive U -> V}| &&
    \coq|f| is an additive function ($\mathbb{Z}$-module homomorphism) from \coq|U| to \coq|V|. \\
  $\begin{array}{@{}l@{}} \coq|U, V : zmodType,| \\ \qquad\coq|f : {additive U -> V}| \end{array}$ &
    \coq|#$\vdash$# f : U -> V| &
    \coq|#$\coloneqq$# #\color{gray}\texttt{@}#GRing.Additive.#apply# #\color{gray}\texttt{U V \_}# f| &
    the underlying function of an additive function (implicit coercion). \\
  \coq{U, V : Type, f : U -> V} &
    $\vdash \begin{array}{@{}l@{}} \coq|[additive of f] :| \\ \qquad\coq|{additive U -> V}| \end{array}$ &&
    the canonical additive function of \coq|f| if it exists; otherwise, it does not type-check. \\
  \hline
  & \coq|#$\vdash$# R : ringType| &
    (RHS) \coq|#$\coloneqq$# GRing.Ring.type| &
    \coq|R| is a (not necessarily commutative) ring. \\
  %
  %
  \coq|R : ringType| & \coq|#$\vdash$# 1 : U| &
    \coq|#$\coloneqq$# GRing.one R| &
    the one (multiplicative identity) of \coq|R|. \\
  \coq|R : ringType, x, y : R| & \coq|#$\vdash$# x * y : R| &
    \coq|#$\coloneqq$# #\color{gray}\texttt{@}#GRing.mul #\color{gray}\texttt{R}# x y| &
    the product of \coq|x| and \coq|y| in \coq|R|. \\
  \coq|R : ringType, n : nat| & \coq|#$\vdash$# n
    \coq|#$\coloneqq$# 1 *+ n| &
    the ring image of \coq|(n : nat)| in \coq|R|. \\
  \coq|R : ringType, n : int| & \coq|#$\vdash$# n
    \coq|#$\coloneqq$# 1 *~ n| &
    the ring image of \coq|(n : int)| in \coq|R|. \\
  $\begin{array}{@{}l@{}} \coq|R : ringType,| \\ \qquad\coq|x : R, n : int| \end{array}$ &
    \coq|#$\vdash$# x ^+ n : R| &
    \coq|#$\coloneqq$# #\color{gray}\texttt{@}#GRing.exp #\color{gray}\texttt{R}# x n| &
    the \coq|n|-th power of \coq|x| in \coq|R|. \\
  \hline\pagebreak\hline
  \coq|R, S : ringType| & \coq|#$\vdash$# f : {rmorphism R -> S}| &&
    \coq|f| is a ring homomorphism from \coq|R| to \coq|S|. \\
  $\begin{array}{@{}l@{}} \coq|R, S : ringType,| \\ \qquad\coq|f : {rmorphism R -> S}| \end{array}$ &
    \coq|#$\vdash$# f : R -> S| &
    \coq|#$\coloneqq$# #\color{gray}\texttt{@}#GRing.RMorphism.#apply# #\color{gray}\texttt{R S \_}# f| &
    the underlying function of a ring homomorphism (implicit coercion). \\
  \coq{R, S : Type, f : R -> S} &
    $\vdash \begin{array}{@{}l@{}} \coq|[rmorphism of f] :| \\ \qquad\coq|{rmorphism R -> S}| \end{array}$ &&
    the canonical ring homomorphism of \coq|f| if it exists; otherwise, it does not type-check. \\
  \hline
  & \coq|#$\vdash$# R : comRingType| &
    (RHS) \coq|#$\coloneqq$# GRing.ComRing.type| &
    \coq|R| is a ring whose multiplication is commutative. \\
  %
  \hline
  & \coq|#$\vdash$# R : unitRingType| &
    (RHS) \coq|#$\coloneqq$# GRing.UnitRing.type| &
    \coq|R| is a ring with computable inverses. \\
  %
  %
  \coq|R : unitRingType, x : R| &
    \coq|#$\vdash$# x^-1 : R| &
    \coq|#$\coloneqq$# #\color{gray}\texttt{@}#GRing.inv #\color{gray}\texttt{R}# x| &
    the multiplicative inverse of \coq|x| if exists; otherwise \coq|x| itself. \\
  \coq|R : unitRingType, x, y : R| &
    \coq|#$\vdash$# x / y : R| &
    \coq|#$\coloneqq$# x * y^-1| &
    \coq|x| divided by \coq|y|. \\
  $\begin{array}{@{}l@{}} \coq|R : unitRingType,| \\ \qquad\coq|x : R, n : nat| \end{array}$ &
    \coq|#$\vdash$# x ^- n : R| &
    \coq|#$\coloneqq$# (x ^+ n)^-1| &
    the inverse of \coq|(x ^+ n)|. \\
  $\begin{array}{@{}l@{}} \coq|R : unitRingType,| \\ \qquad\coq|x : R, n : int| \end{array}$ &
    \coq|#$\vdash$# x ^ n : R| &
    \coq|#$\coloneqq$# #\color{gray}\texttt{@}#exprz #\color{gray}\texttt{R}# x n| &
    the \coq|n|-th power of \coq|x|,
    reduces to either \coq|(x ^+ _)| or \coq|(x ^- _)|. \\
  \hline
  & \coq|#$\vdash$# R : comUnitRingType| &
    (RHS) \coq|#$\coloneqq$# GRing.ComUnitRing.type| &
    \coq|R| is a commutative ring with computable inverses. \\
  %
  \hline
  & \coq|#$\vdash$# F : fieldType| &
    (RHS) \coq|#$\coloneqq$# GRing.Field.type| &
    \coq|F| is a field. \\
  %
  \hline
  & \coq|#$\vdash$# R : numDomainType| &
    (RHS) \coq|#$\coloneqq$# Num.NumDomain.type| &
    \coq|R| is a partially ordered integral domain. \\
  %
  \hline
  & \coq|#$\vdash$# F : numFieldType| &
    (RHS) \coq|#$\coloneqq$# Num.NumField.type| &
    \coq|F| is a field with a partial order and a norm. \\
  %
  \hline
 \end{longtable}
\end{landscape}

\end{document}

%% file: macro.tex
\newminted[elpicode]{elpi.py:ElpiLexer -x}{fontsize=\small,linenos=true,escapeinside=\#\#,texcomments}
\newmintinline[elpi]{elpi.py:ElpiLexer -x}{escapeinside=\#\#}
\newminted[coqcode]{elpi.py:CoqElpiLexer -x}{fontsize=\small,linenos=true,escapeinside=\#\#,texcomments}
\newmintinline[coq]{elpi.py:CoqElpiLexer -x}{escapeinside=\#\#}

\tikzset{
every node/.style={inner sep=2},
every path/.style={thick},
every picture/.style={font issue=\footnotesize},
font issue/.style={execute at begin picture={#1\selectfont}},
strike through/.append style={
    decoration={markings, mark=at position 0.5 with {
    \draw[-] ++ (-2pt,-2pt) -- (2pt,2pt);}
  },postaction={decorate}},
structure/.style={}
}

\newcommand\conveq{\equiv}
\newcommand\unify{\ensuremath{\mathrel{\hat{\conveq}}}}

\newcommand{\Coq}{{\sffamily Coq}}

\newcommand{\Ltac}{{\sffamily Ltac}}
\newcommand{\Ltactwo}{{\sffamily Ltac2}}
\newcommand{\Mtactwo}{{\sffamily Mtac2}}
\newcommand{\CoqElpi}{{\sffamily Coq-Elpi}}
\newcommand{\CoqEAL}{{\sffamily CoqEAL}}
\newcommand{\Elpi}{{\sffamily Elpi}}
\newcommand{\ssr}{{\sffamily SSReflect}}
\newcommand{\MC}{{\sffamily MathComp}}
\newcommand{\Agda}{{\sffamily Agda}}
\newcommand{\Lean}{{\sffamily Lean}}

\newcommand{\sect}{Section~}
\newcommand{\chap}{Chapter~}
\newcommand{\appx}{Appendix~}
\newcommand{\fig}{Figure~}